\documentclass[9pt,twocolumn,twoside]{opticajnl}
\journal{opticajournal} 

\setboolean{shortarticle}{false}

\usepackage{lineno}

\usepackage[svgnames,table]{xcolor}

\usepackage{braket}

\begin{document}

\title{Demonstration of a next-generation wavefront actuator for gravitational-wave detection}

\author[1]{Tyler Rosauer}
\author[2]{Huy Tuong Cao}
\author[1]{Mohak~Bhattacharya}
\author[1]{Peter~Carney}
\author[1]{Luke~Johnson}
\author[1]{Shane~Levin}
\author[1]{Cynthia~Liang}
\author[1]{Xuesi~Ma}
\author[1]{Luis~Martin~Gutierrez}
\author[1]{Michael~Padilla}
\author[1]{Liu~Tao}
\author[1]{Aiden~Wilkin}
\author[3]{Aidan~Brooks}
\author[1,*]{Jonathan W. Richardson}

\affil[1]{Department of Physics \& Astronomy, University of California, Riverside, Riverside, CA 92521, USA}
\affil[2]{LIGO Laboratory, Massachusetts Institute of Technology, Cambridge, MA 02139, USA}
\affil[3]{LIGO Laboratory, California Institute of Technology, Pasadena, CA 91125, USA}
\affil[*]{jonathan.richardson@ucr.edu}

\begin{abstract}
In the last decade, the Laser Interferometer Gravitational-Wave Observatory (LIGO) and the European Virgo observatory have opened a new observational window on the universe. These cavity-enhanced laser interferometers sense spacetime strain, generated by distant astrophysical events such as black hole mergers, to an RMS fluctuation of a few parts in $10^{21}$ over a multi-kilometer baseline. Optical advancements in laser wavefront control are key to advancing the sensitivity of current detectors and enabling a planned next-generation 40-km gravitational wave observatory in the United States, known as Cosmic Explorer. We report the first experimental demonstration of a new wavefront control technique for gravitational-wave detection, obtained from testing a full-scale prototype on a 40-kg LIGO mirror. Our results indicate that this design can meet the unique and challenging requirements of providing higher-order precision wavefront corrections at megawatt laser power levels, while introducing extremely low effective displacement noise into the interferometer. This new technology will have a direct and enabling impact on the observational science, expanding the gravitational-wave detection horizon to very early times in the universe, before the first stars formed, and enabling new tests of gravity, cosmology, and dense nuclear matter.
\end{abstract}

\setboolean{displaycopyright}{false} 

\maketitle


\section{Introduction}
\label{sec:intro}
Nine years after the Laser Interferometer Gravitational-Wave Observatory (LIGO) first detected gravitational waves from two coalescing black holes~\cite{GW150914}, opening a new window on the universe, design work is starting on a new generation of detectors capable of enabling a much broader range of observational science. Two successive upgrades of the 4-km LIGO detectors, A+~\cite{Aplus} followed by $\rm A^{\#}$~\cite{PostO5Report:2022}, will pave the way to Cosmic~Explorer~\cite{CEHorizonStudy}, a next-generation 40-km gravitational-wave observatory in the U.S. with ten times greater sensitivity than current detectors. Cosmic~Explorer will expand the gravitational-wave detection horizon to cosmological redshifts of up to~100, corresponding to a time before the first stars formed, when the universe was only 0.1\% of its present age. It will observe hundreds of thousands to millions of binary black hole and neutron star mergers per year across cosmic time.

Accessing this large compact binary population will enable measurement of the expansion history of the universe to sub-percent precision~\cite{Chen:2018}, across a redshift range that is complementary to cosmic microwave background and local distance ladder measurements~\cite{Farr:2019} and may resolve the growing tension between them~\cite{Riess:2022, Murakami:2023}. Local black hole mergers will be detected with signal-to-noise ratios in excess of 1000 (compared to 10-26 currently), enabling precision tests of strong-field general relativity and the nature of black holes~\cite{Skenderis:2008, Cardoso:2017, Brustein:2018}. Binary neutron stars will be detected \textit{minutes} before merger, enabling multi-messenger partners to pre-point their instruments and observe potential precursor events, such as flares due to neutron crust shattering~\cite{Tsang:2012}.

Across most of the gravitational-wave frequency band accessible to ground-based detectors, quantum noise is the limiting source of instrumental noise. It arises from the quantization of the electromagnetic field used to probe the positions of the interferometer's mirrors~\cite{Caves:1980, Caves:1981}. Below 100~Hz, amplitude-quadrature fluctuations of the optical field, which physically displace the mirrors via radiation pressure, dominate the quantum noise. At higher frequencies, phase-quadrature fluctuations, manifesting as shot noise, dominate the quantum noise. Higher laser power reduces the amplitude spectral density of the shot noise as $1/\sqrt{P_{\rm arm}}$, where $P_{\rm arm}$ is the interferometer arm power. Additionally, the injection of a frequency-dependent squeezed vacuum field~\cite{Barsotti:2018, Tse:2019} reduces the broadband amplitude spectral density of the quantum noise as $10^{-r/20}$, where $r$ is the effective squeezing factor in decibels.

In future gravitational-wave detectors, the achievable laser power and quantum squeezing will be limited by our ability to correct thermally-induced aberrations in the interferometer's core optics, known as ''test masses.'' Based on experience with the Advanced~LIGO optics, each coated optical surface is expected to absorb roughly 0.5~ppm of incident laser power~\cite{Brooks:2016}. Absorption creates temperature gradients within the test masses that produce optical aberrations through (1)~thermoelastic deformation of the optical surfaces and (2)~thermorefractive lensing within the substrate~\cite{Brooks:2016}. These thermal distortions increase in magnitude proportionally to the incident laser power. They result in excess optical loss as well as coherent mode-scattering effects that degrade both the power buildup and the squeezing, directly increasing the noise floor of the detector~\cite{Buikema:2020, Brooks:2021, McCuller:2021}. Thus, thermal distortions will impose the practical limit on the quantum-limited sensitivity of future detectors.

\begin{figure}[t]
    \centering
    \includegraphics[width=1\linewidth]{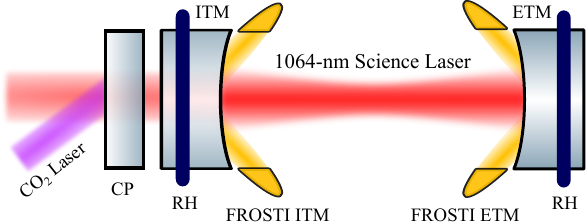}
    \caption{LIGO's thermal compensation system (TCS), shown for one 4-km arm cavity (the other arm is identical). The functions of the ring heaters (RH) encircling each test mass (ITM/ETM) and the compensation plate (CP) are described in the text and in further detail in Ref.~\cite{Brooks:2016}. The FROSTI devices (yellow) are a new front-surface wavefront actuator proposed in this work.}
    \label{fig:tcs_topology}
\end{figure}

In order to achieve their sensitivity targets, LIGO $\rm A^{\#}$ and Cosmic Explorer will each require an unprecedented 1.5~MW of circulating laser power in their arm cavities and 10~dB of squeezed-light-enabled noise reduction. This is roughly a fourfold increase beyond the highest levels achieved in LIGO today~\cite{Buikema:2020}. Modeling has shown that this will require a {\it qualitatively} new form of active wavefront correction on the front surface of the test masses, targeting finer spatial scales than those accessible to LIGO's existing thermal compensation system (TCS)~\cite{AsharpTCSReqs:2022}. Figure~\ref{fig:tcs_topology} shows LIGO's current TCS capabilities. Inside the high-power arm cavities, only the lowest-order aberrations can currently be corrected: tilt, through alignment servos, and defocus, by actuating the radii of curvature of the test masses. Radius of curvature actuation is provided by a ring heater which encircles the barrel of each test mass and radiatively heats the substrate~\cite{Brooks:2016}. An additional compensation plate, located outside each arm cavity, is heated by a 10.6-$\mu$m $\rm CO_2$ laser in order to partially correct the residual substrate lens of the input test mass (ITM)~\cite{Brooks:2016}.

\begin{figure}[t]
    \centering
    \includegraphics[width=\linewidth]{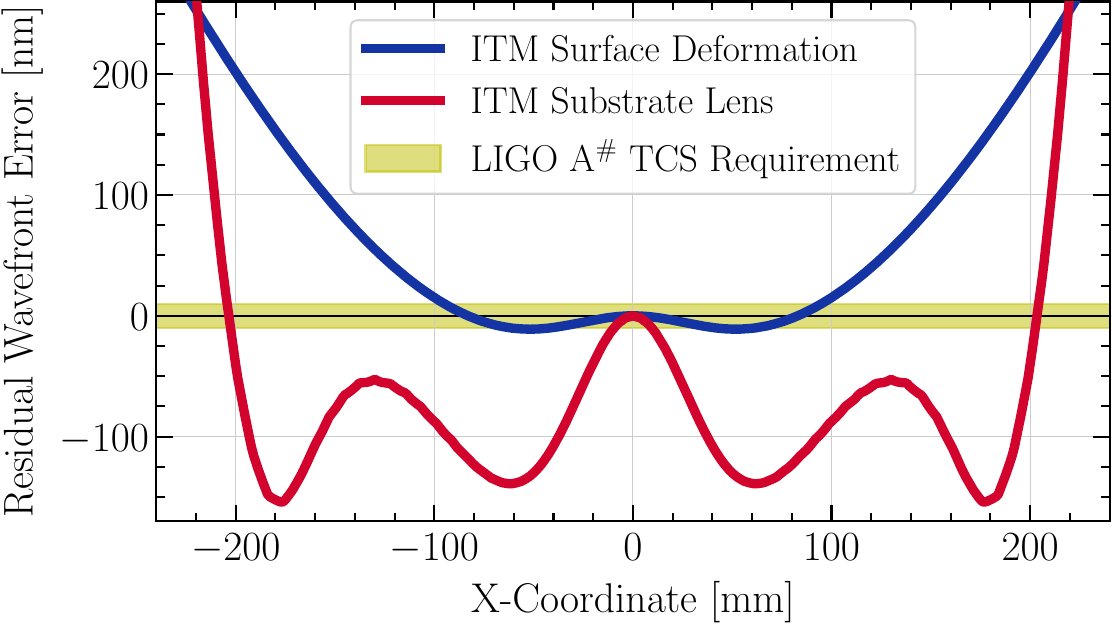}
    \caption{Residual wavefront errors on a 100-kg input test mass (ITM) at 1.5~MW of incident laser power, {\it after} optimal correction with LIGO's existing thermal compensation system (TCS). The thermoelastic surface deformation (blue) and thermorefractive substrate lens (red) are imprinted on the reflected and transmitted laser fields, respectively. In contrast, the yellow band shows the residual error requirement for LIGO~$\rm A^{\#}$~\cite{AsharpTCSReqs:2022}.}
    \label{fig:wavefront_error}
\end{figure}

The LIGO ring heaters produce a purely spherical wavefront actuation which is well matched to the thermal distortion near the center of the optic, but which results in a residual edge rise that becomes larger with increasing incident laser power~\cite{Brooks:2016}. Figure~\ref{fig:wavefront_error}, for example, shows the residual wavefront error introduced by an ITM at 1.5~MW of incident power in LIGO~$\rm A^{\#}$, \textit{after} optimal correction with LIGO's existing TCS actuators. Modeling has shown that these residual wavefront errors must be reduced to less than roughly 10~nm RMS across the full aperture of the test masses~\cite{AsharpTCSReqs:2022}. Thus, a new wavefront control technique is required to apply corrections near the edge of the optic.

In this article, we report the development of a new wavefront actuation technology capable of providing targeted higher-order laser wavefront  corrections in gravitational-wave detectors. It is key to enabling LIGO~$\rm A^{\#}$ and opens a new research and development pathway towards next-generation gravitational-wave observatories. In \S\ref{sec:approach}, we first present the design principles underlying our approach and the unique challenges which drive them. In \S\ref{sec:results}, we then present the experimental results of the testing of a full-scale prototype actuator on a 40-kg LIGO test mass. Finally, in \S\ref{sec:discussion} we discuss the foundation this work lays for realizing a next-generation gravitational-wave observatory, such as Cosmic Explorer.

\section{Approach}
\label{sec:approach}

\begin{figure*}[t]
    \centering
    \includegraphics[width=1\textwidth]{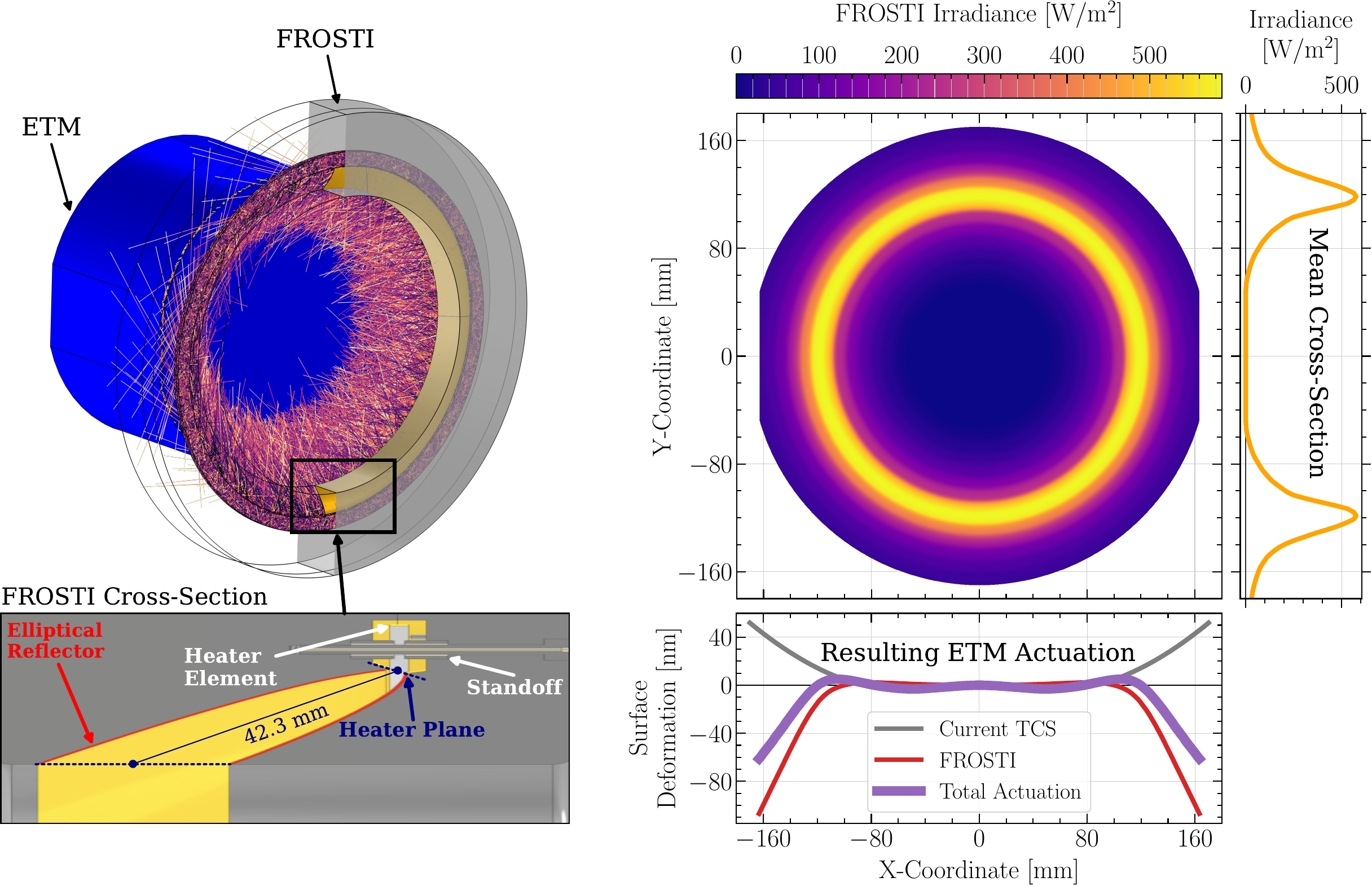}
    \caption{Ray-tracing and finite-element simulations of the FROnt Surface Type Irradiator (FROSTI) designed to apply higher-order wavefront corrections to the LIGO test masses. Each actuator mounts 5~cm in front of the test mass and just outside its 34-cm diameter (top left). It projects an annular radiation pattern onto the front surface of the test mass, where it is absorbed (top right). The resulting surface deformation profile (bottom right) is designed to (1)~more accurately compensate the coating absorption and (2)~eliminate a higher-order mode co-resonance (HOM7) in the arm cavities, by producing extra roll-off near the edge (see Ref.~\cite{Richardson:2022}).}
    \label{fig:frosti_concept}
\end{figure*}

Higher-order active wavefront control in gravitational-wave detectors cannot be achieved with a straightforward extension of current technology. The following sections discuss the key drivers of these requirements and introduce a new approach to meeting them, which has led to the development of a prototype actuator demonstrating the effectiveness of this technique.

\subsection{Key Challenges}
\label{sec:challenges}
Reducing the residual distortion that remains after the laser beam heating is corrected with LIGO's existing ring heaters (see Figure~\ref{fig:wavefront_error}) requires a new form of wavefront actuation on smaller (2-5~cm) spatial scales. Moreover, in order to mitigate thermally-induced losses arising \textit{inside} the high-power arm cavities, there is no viable alternative to applying a corrective heating profile directly to the front (reflective-coated) surface of each test mass. However, projecting a corrective heating pattern onto the test masses, the most displacement-sensitive optics which couple directly to gravitational waves, poses a major experimental challenge. Intensity fluctuations of the incident radiation will displace the test masses through optomechanical and photothermal mechanisms~\cite{Ballmer:2006, Willems:2011}. These effects include both radiation pressure and photothermally-induced expansion and contraction of the optical surface layer, which couple the intensity noise of the wavefront actuators \textit{directly} to the interferometer's readout signal. LIGO's stringent requirements on induced displacement noise exclude the use of traditional heating beam sources, such as 10.6-$\mu\mathrm{m}$ $\mathrm{CO}_2$ lasers, and are described further in \S\ref{sec:rin}.

\subsection{FROnt Surface Type Irradiator (FROSTI) Concept}
\label{sec:frosti}

\begin{figure*}[t]
    \centering
    \includegraphics[width=0.327\linewidth, trim=21.4cm 4.7cm 3cm 10.3cm, clip=true]{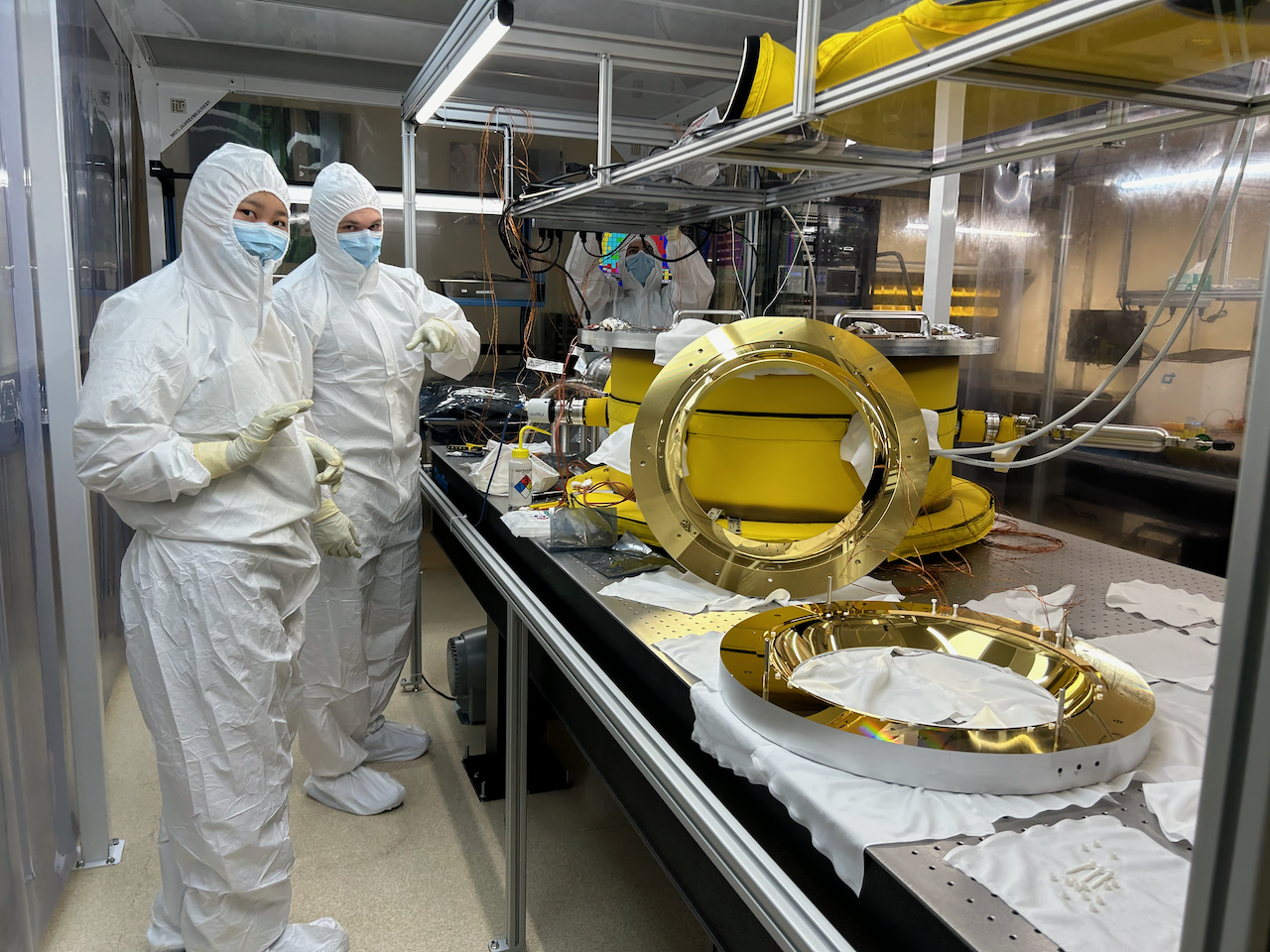}
    \hspace{0.15cm}
    \includegraphics[width=0.385\linewidth, trim=0.45cm 0cm 0cm 0cm, clip=true]{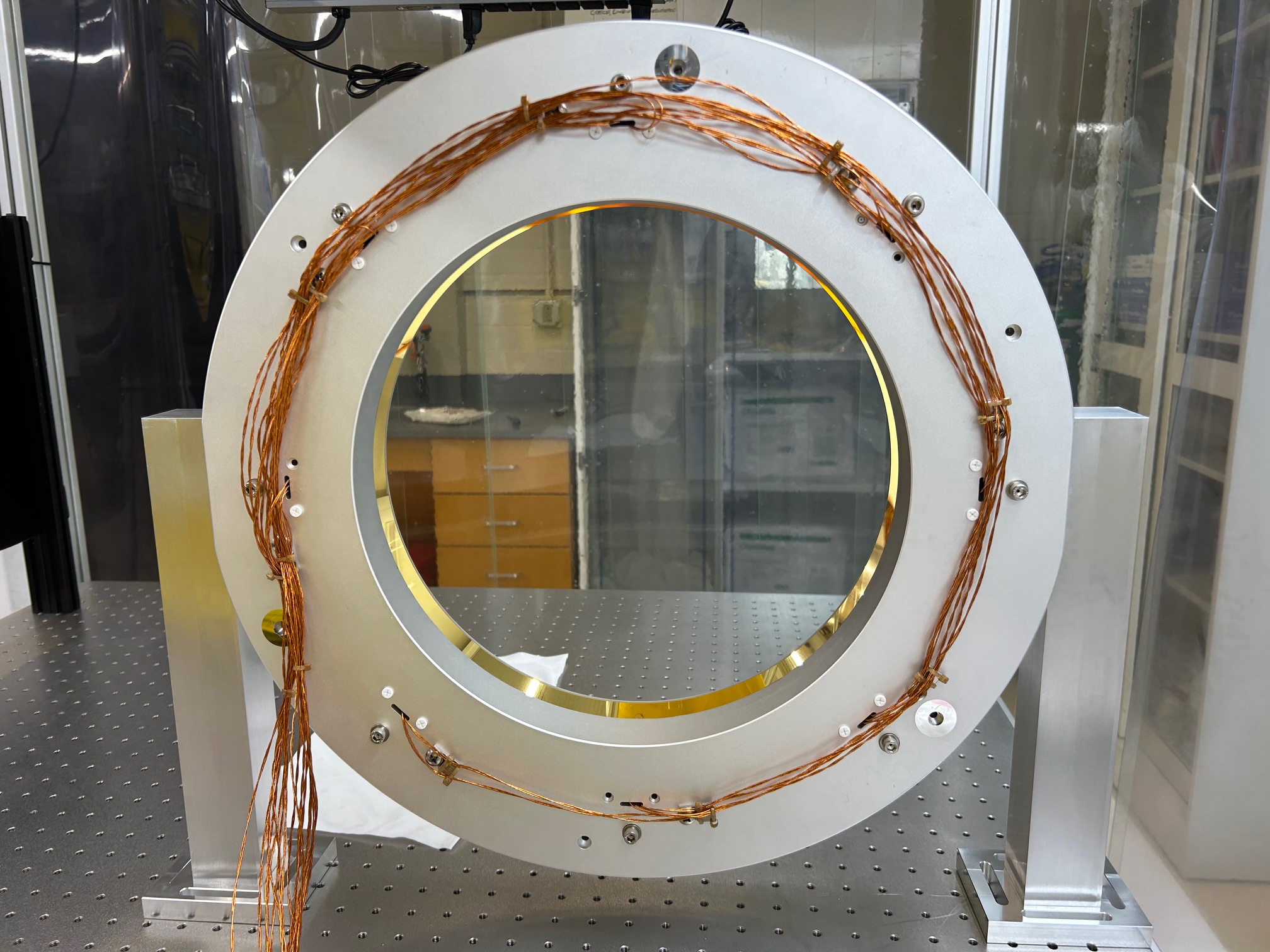}
    \hspace{0.15cm}
    \includegraphics[width=0.254\linewidth, trim=4cm 0.9cm 3.5cm 0.7cm, clip=true]{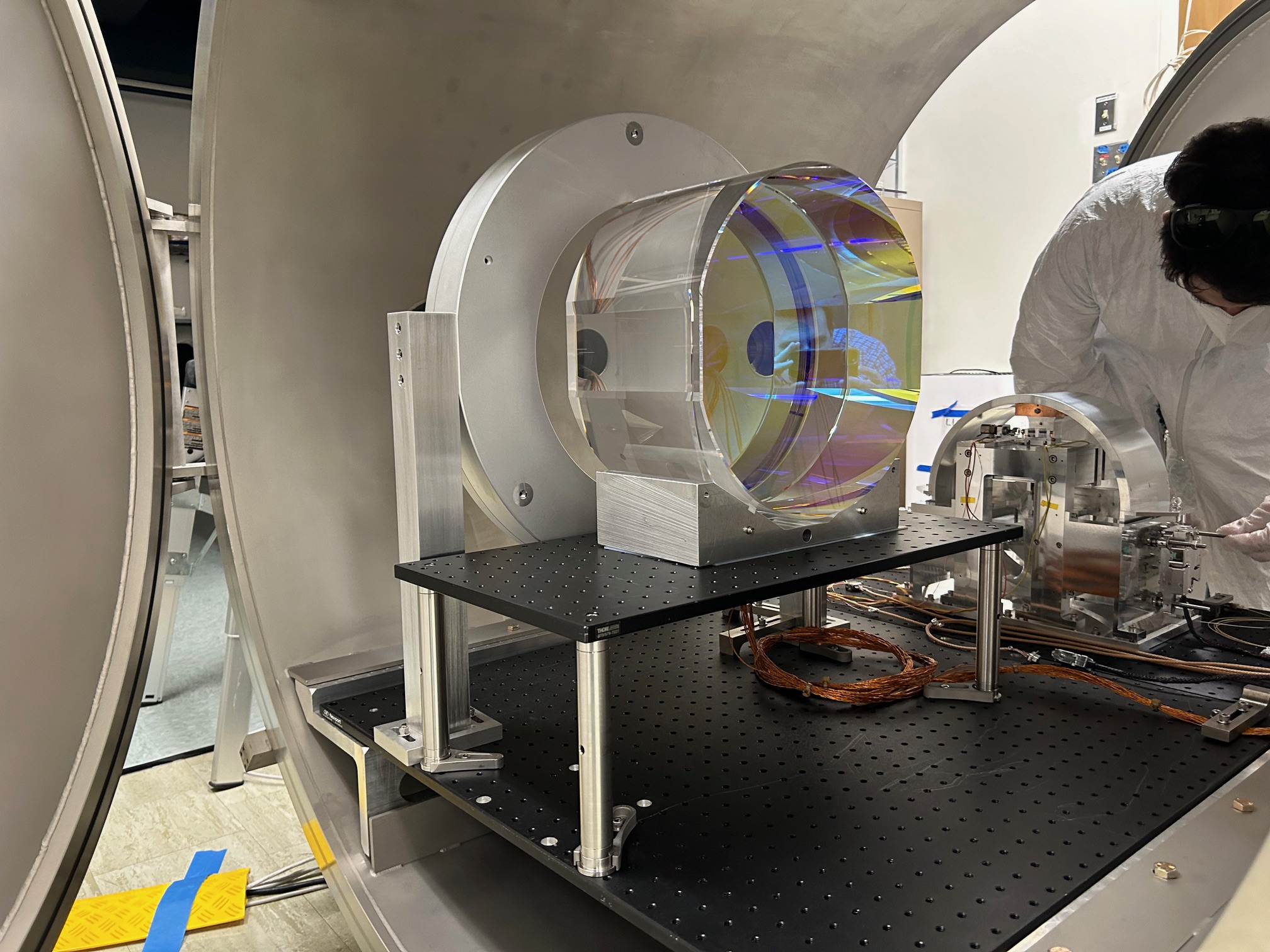}
    \caption{Photos of the FROnt Surface Type Irradiator (FROSTI) prototype, fabricated to scale for actuation on a 40-kg LIGO test mass. \textit{Left:} The two ``upper'' and ''lower'' reflector sections, which are joined to form an annular nonimaging elliptical trough. \textit{Center:} The fully-assembled FROSTI prototype, looking towards the rear surface which faces away from the test mass. \textit{Right:} Test configuration showing the positioning of the FROSTI (left) relative to the test mass (center), whose  reflective surface is facing toward the FROSTI.}
    \label{fig:prototype}
\end{figure*}

The FROnt Surface Type Irradiator (FROSTI) is our concept for a new generation of wavefront actuators that can provide low-noise wavefront control on few-centimeter spatial scales. FROSTI is named for its function of restoring a test mass to its cold {\it optical} state (thermoelastically and thermorefractively) while illuminated by megawatt laser power. Each actuator mounts in vacuum approximately 5~cm from the reflective front surface of the test mass, as shown in Figure~\ref{fig:frosti_concept} (top left), and just outside the 34-cm coated diameter. Their principle of operation is to project 3-14~$\mu$m thermal radiation, produced by an internal greybody source and reshaped by nonimaging reflectors into an annular spatial pattern, onto the front surface of each test mass.

The radiation pattern can be optimized to induce the ideal thermoelastic and thermorefractive responses for each optic. For example, Figure~\ref{fig:frosti_concept} (top middle) shows the optimal irradiance profile for a 40-kg end test mass (ETM). As is visible in its one-dimensional cross-section (top right), the irradiance profile has a long tail which is small (containing less than 0.1\% of the radiated power) but nonzero at the edge of the optic, due to unconstrained azimuthal rays.
The resulting surface deformation (bottom right) is designed to more accurately correct the laser beam heating, near the center of the optic, while also mitigating a higher-order mode co-resonance in the arm cavities, by producing extra roll-off near the edge (see Ref.~\cite{Richardson:2022}).

Thermal radiation is currently the only source with sufficient intensity stability to meet LIGO's stringent noise requirements. Blackbody or greybody sources can radiate with an intensity stability approaching the shot noise limit for a given level of emitted power. Strictly, the intensity noise of a blackbody source is larger than the shot noise of a non-thermal source of equivalent power. Since thermally-emitted photons, described by Bose-Einstein statistics, arrive not randomly in time but in bunches, the root-mean-square (RMS) fluctuation in their arrival rate is larger by a factor of 
\begin{equation}
    \frac{\sigma_{\mathrm{thermal}}}{\sigma_{\mathrm{shot}}} = \sqrt{\frac{e^{h\nu/kT}}{e^{h\nu/kT} - 1}}
\end{equation}
relative to shot noise of the same average emitted power~\cite{VanValkenburg:2001}. However, near the peak emission wavelength of a realistic 400~C source (4.3~$\mu$m), the RMS fluctuation is larger by less than 0.4\%. Thus this distinction is insignificant from a practical perspective.

While the FROSTI concept pushes the technology into a new regime, thermal photon sources have already proven effective for low-noise active wavefront control in gravitational-wave detectors. In the Virgo+ detector, the radii of curvature of the end test masses were controlled using radiation from a high-temperature greybody emitter mounted inside an elliptical reflector~\cite{Accadia:2013}. This actuator, known as the CHRoCC (Central Heating Radius of Curvature Correction), was mounted in vacuum to the seismic isolation platform and projected its radiation upwards onto the front surface of the test mass. In the GEO~600 detector, a multi-heater array, known as the Matrix Heater, has been used to compensate the beamsplitter from outside the vacuum system with high spatial resolution~\cite{Wittel:2018}. However, its power delivery efficiency is limited to only 1\% percent by the numerical aperture of the vacuum viewport, in comparison to an estimated 56\% delivery efficiency for the in-vacuum CHRoCC. Due to the significantly higher power delivery requirement for test mass compensation (up to 25~W in the $\rm A^{\#}$ era), an out-of-vacuum actuator was judged to be impractical for use in LIGO.

\subsection{Prototype Design and Fabrication}
\label{sec:prototype}
To demonstrate the FROSTI concept, a full-scale prototype of the wavefront actuator depicted in Figure~\ref{fig:frosti_concept} was fabricated. Its irradiance profile, designed for a 40-kg LIGO ETM, is optimized through finite element analysis (FEA) to
\begin{enumerate}
    \item Reduce the residual, high-spatial-frequency surface deformation due to coating absorption (see Figure~\ref{fig:wavefront_error}, blue curve).
    \item Eliminate higher-order mode co-resonances in LIGO's arm cavities, using additional edge roll-off to shift their resonant frequencies \textit{relative to} the fundamental mode (see Ref.~\cite{Richardson:2022}).
\end{enumerate}
Figure~\ref{fig:frosti_concept} (bottom right) shows the surface profile actuation that results from applying this irradiance profile to the ETM. Although the residual substrate lens (shown in Figure~\ref{fig:wavefront_error}, red curve) will be an equally important design driver of an irradiance profile for the partially-transmissive ITMs, it is not a relevant design consideration for the highly-reflective ETMs because their transmitted beam path is not a part of the interferometer.

The annular irradiance pattern is produced by a nonimaging reflector which encases the internal greybody radiation source. The reflector is formed from two asymmetric elliptical curves, shown in Figure~\ref{fig:frosti_concept} (bottom left), which are revolved $360^{\circ}$ to form a three-dimensional elliptical trough. Each elliptical curve is designed using the nonimaging edge-ray technique~\cite{Winston:2005, Jiang:2015}, which aims to achieve maximum delivery efficiency of radiation from a source plane of finite dimension (the greybody emitter) to a target plane (the test mass surface). Ray-tracing analysis indicates that our design achieves a power delivery efficiency of 85\% to the target region of the test mass surface. A detailed description of our design procedure, and the resulting elliptical curve parameters, are presented in the Supplemental Document. The prototype's reflector consists of two sections, each containing one of the two optical surfaces, which are bolted together. Both are produced by NiPro Optics of Irvine, CA. Each section is fabricated from 6061~aluminum via single-point diamond turning, a computerized machining technique which naturally achieves a 5~nm~RMS surface finish. A gold thin-film coating is deposited onto each reflector surface to achieve an enhanced reflectivity greater than 98\% across the infrared band. The two fabricated reflector sections are pictured in Figure~\ref{fig:prototype} (left).

The greybody radiation source is a 400~mm-diameter heater ring consisting of aluminum nitride, an ultra high vacuum (UHV) compatible ceramic, with embedded tungsten wiring for power as well as temperature sensing via an integrated resistance temperature detector (RTD). Eight individual heater elements, produced by Oasis Materials of Poway, CA, are mounted end-to-end to form the ring. In order to ensure that the heater elements predominantly radiate through their front surfaces, which face outward into the reflector, a gold thin-film coating is deposited onto all of the interior-facing surfaces. This lowers their emissivity from 0.90 to approximately 0.02 across the infrared band. Moreover, in order to minimize conductive thermal losses, Macor ceramic standoffs are used to prevent the heater elements from physically contacting the aluminum reflector. At the heater elements' maximum operating temperature of 400~C, they are designed to deliver 25~W to the test mass surface. Figure~\ref{fig:prototype} (center) shows the fully assembled FROSTI prototype.

\section{Experimental Results}
\label{sec:results}
The FROSTI prototype testing procedures are designed to confirm the following:
\begin{enumerate}
    \item Measured surface temperature and wavefront actuation profiles on a real LIGO ETM consistent with the design targets shown in Figure~\ref{fig:frosti_concept}.
    \item Noise spectra that meet the stringent requirements for the LIGO A+ and $\rm A^{\#}$ upgrades.
    \item Ultra high vacuum (UHV) environment compatibility.
\end{enumerate}
To accomplish the above, the FROSTI prototype has undergone two phases of testing: in-vacuum tests confirming the optical performance and UHV compatibility, followed by an in-air test measuring the relative intensity noise (RIN) of the device.

\subsection{Optical Performance\label{sec:irradiance}}

\begin{figure}[t]
    \centering
    \includegraphics[width=1\linewidth, trim=0.48cm 0.43cm 1.17cm 0.43cm, clip=true]{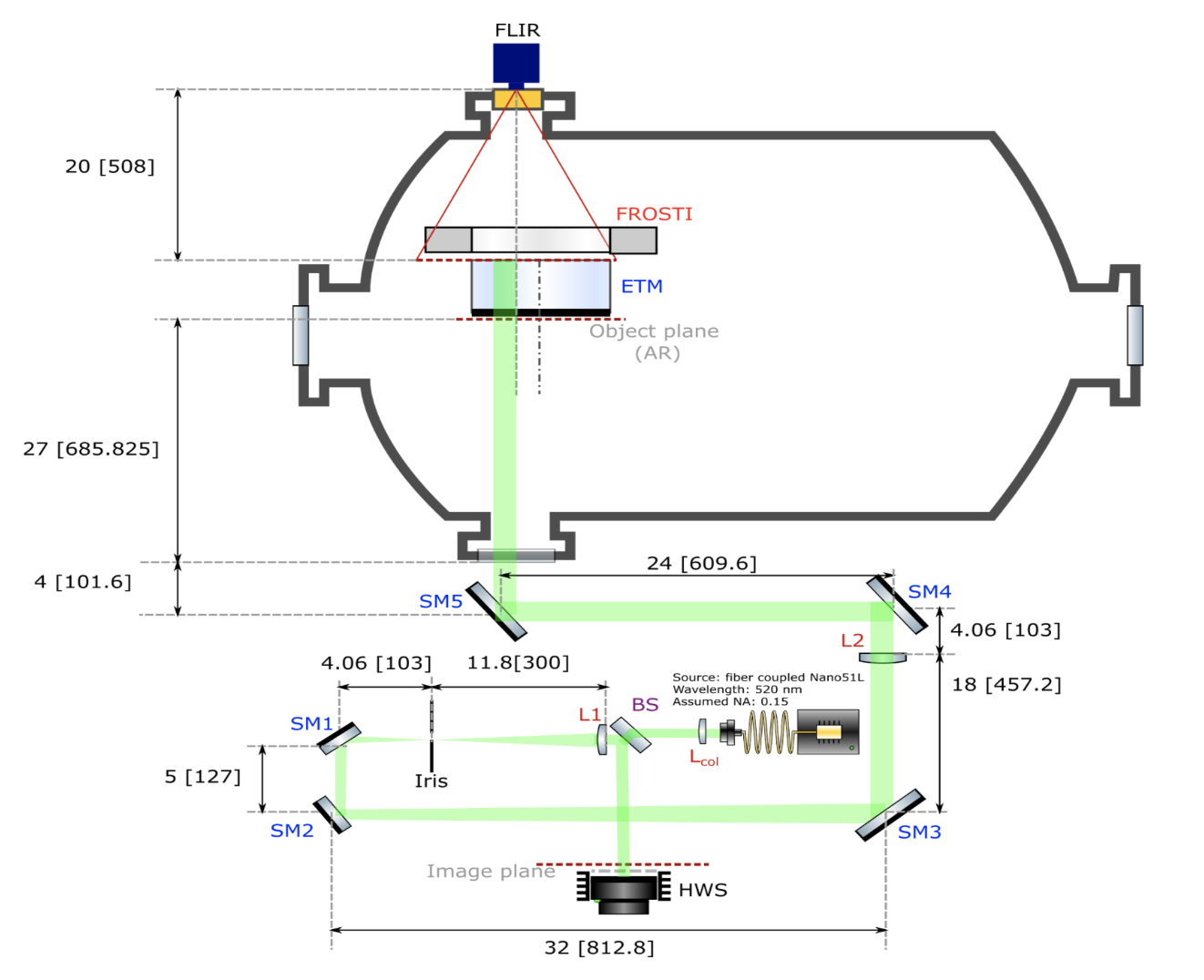}
    \caption{Experimental configuration used to test the FROSTI prototype in vacuum on a 40-kg LIGO ETM. The thermoelastic and thermorefractive responses of the ETM to the FROSTI heating profile are measured using a Hartmann wavefront sensor (HWS) and thermal imaging camera (FLIR).}
    \label{fig:experimental_layout}
\end{figure}

The configuration used to measure the surface temperature and wavefront actuation profiles on an ETM is shown in Figure~\ref{fig:experimental_layout}. The FROSTI prototype is placed 5~cm in front of the highly-reflective (HR) surface of the ETM, as pictured in Figure~\ref{fig:prototype} (right). The entire HR surface is viewed face-on through a zinc selenide viewport by a FLIR A70 thermal imaging camera. From the opposite side, a 520-nm laser probe beam, injected through another viewport, propagates through the ETM substrate and reflects from the HR surface. A Hartmann wavefront sensor measures the thermoelastic and thermorefractive effects imprinted on the probe beam as the ETM is heated. When heated from room temperature, we found that the ETM takes roughly 10~hours to reach thermal steady-state behavior, consistent with FEA modeling.

\subsubsection{Surface Temperature Variation}

\begin{figure}[t]
    \centering
    \includegraphics[width=1\linewidth]{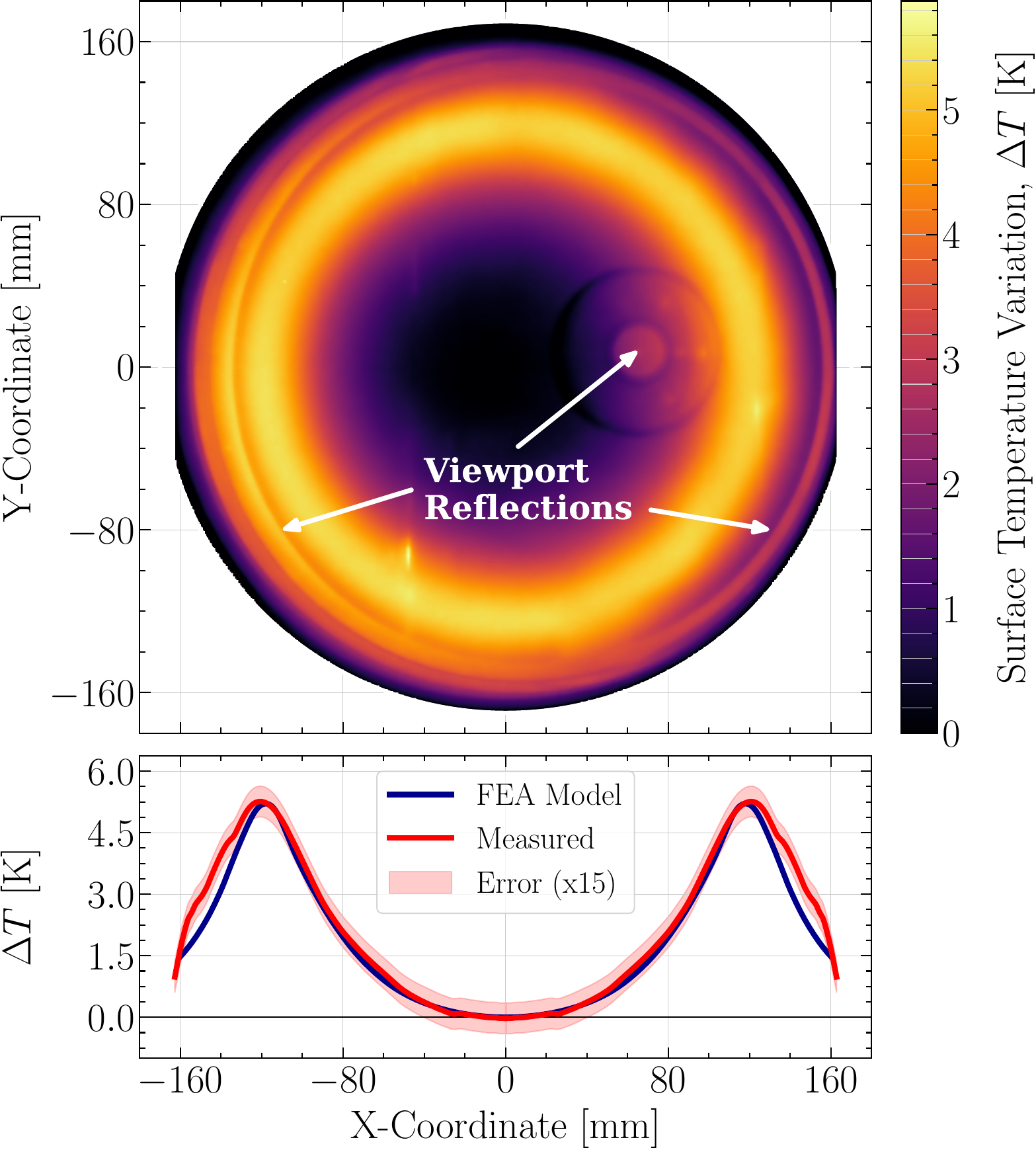}
    \caption{\textit{Top:} Measured surface temperature map of the ETM due to the irradiance profile applied by the FROSTI prototype. The values are expressed as temperature differences relative to the center of the optic, with known measurement artifacts noted. \textit{Bottom:} Radial average of the surface temperature map, shown in comparison to a finite-element model prediction.}
    \label{fig:FROSTI_dT}
\end{figure}

\begin{figure*}[t]
    \centering
    \includegraphics[width=1\linewidth]{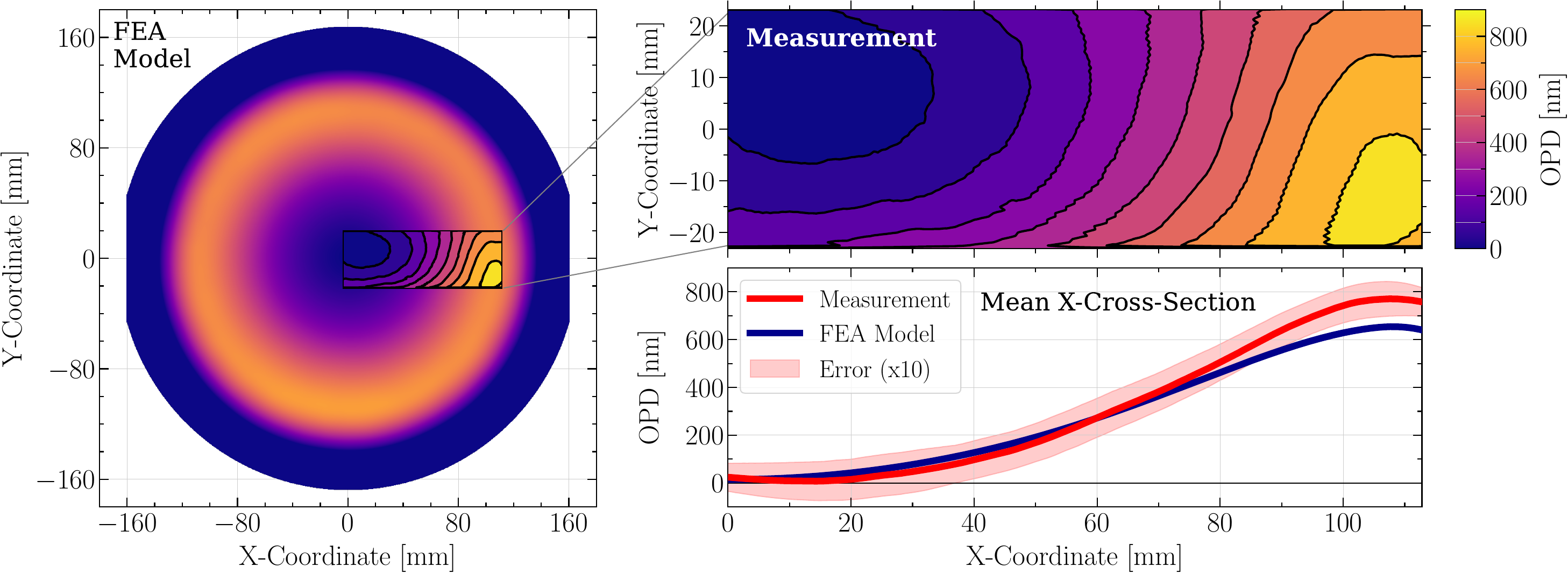}
    \caption{\textit{Left:} FEA-generated map of the optical path difference (OPD) produced by the FROSTI irradiance profile shown in Figure~\ref{fig:frosti_concept}, overlaid with the measurement region of the Hartmann wavefront sensor. \textit{Right:} Measured OPD map, constructed by horizontally translating the Hartmann wavefront sensor's probe beam across the measurement region. Shown below the map is the averaged one-dimensional profile of the OPD in this region, compared to an FEA model prediction with 10.2~W of absorbed FROSTI power.}
    \label{fig:HWS}
\end{figure*}

After balancing the power radiated by the eight individual heater elements and allowing the ETM to reach a thermal steady state, the surface temperature profile of the ETM was measured over the course of 105 minutes using the FLIR camera. Measurements were recorded at one-minute intervals, giving 106 samples of the surface temperature in total. These samples are averaged to obtain the surface temperature map shown in Figure~\ref{fig:FROSTI_dT} (top), whose values are expressed as temperature \textit{differences} relative to the center of the optic. The corresponding radial average of this temperature map is shown in Figure~\ref{fig:FROSTI_dT} (bottom). 

Several aberrations visible in the surface temperature map are known measurement artifacts, as annotated in Figure~\ref{fig:FROSTI_dT} (top). These artifacts arise from reflections between the ETM's HR surface and the zinc selenide viewport used by the thermal imaging camera (see Figure~\ref{fig:experimental_layout}), which directly faces the ETM and lacks an anti-reflection coating. Although the reflectivity of the LIGO HR coating has not been measured at wavelengths beyond 1064~nm, we are able to constrain its effective broadband infrared reflectivity, at a $45^{\circ}$ angle of incidence, to be approximately 0.15 by matching the surface temperature and wavefront sensor measurements to a \textit{joint} FEA model of both effects.

The FEA model assumes an irradiance profile proportional to the design profile shown in Figure~\ref{fig:frosti_concept} (top middle), which is incident on a 40-kg ETM in vacuum. Some fraction of the incident power is reflected by the HR coating, while the remainder transmits through the coating and is absorbed in the surface layer of the fused silica substrate. The material properties of the low-OH fused silica used for LIGO's test masses, Heraeus Suprasil 3001, are listed in Section~2 of the Supplemental Document. The ETM is assumed to initially be at 298~K, the measured temperature of its enclosing vacuum environment. We then compute the steady-state heat transfer solution, for the applied FROSTI heat flux, to obtain the resulting surface temperature and wavefront actuation maps. We find that the best-fit model most closely reproducing the FLIR and wavefront sensor measurements corresponds to 12.0~W of incident power, of which 10.2~W is absorbed. This results in a peak temperature difference of 5.21~K between the ETM's center and outer radii, which is in close agreement with the measured value of $5.26 \pm 0.03$~K.

\subsubsection{Wavefront Actuation}
In addition to the full-surface temperature measurement, the wavefront actuation produced by the applied FROSTI heating pattern is directly measured, across a smaller region of the ETM, using a Hartmann wavefront sensor (see Figure~\ref{fig:experimental_layout}). The measured optical path difference (OPD) is obtained by translating the 520-nm probe beam along a horizontal section of the ETM's HR surface, bounded to the left by the center of the mirror and extending rightwards to a maximum radial distance of 113~mm. The translated samples are then combined to reconstruct the total induced OPD across the measurement region, as shown in the top right panel of Figure~\ref{fig:HWS}. The corresponding one-dimensional average of the OPD in this region is shown in the bottom right panel. We find that it is in reasonable agreement with the best-fit FEA model described in the previous section. Both curves attain their maximum value at the location of the applied FROSTI irradiance, with a measured peak OPD of $771 \pm 7$~nm compared to a modeled peak OPD of 654~nm. This measurement directly demonstrates the FROSTI's capability to mitigate residual wavefront distortions at the outer radii of the LIGO test masses.

\subsection{Noise Performance}
\label{sec:noise}
There are several pathways through which FROSTI actuators can inject noise into the interferometer. The two primary sources of noise are intensity noise, originating from power fluctuations of the applied heating profile, and backscattered light noise, due to scattered 1064-nm laser light reflecting from the FROSTI back into the main beam path. The sum of all equivalent displacement noises due to FROSTI actuators must be at least a factor of ten smaller than the design sensitivity of the LIGO A+ detectors (in units of amplitude spectral density) at all frequencies.

\subsubsection{Relative Intensity Noise}
\label{sec:rin}
Relative intensity noise (RIN), due to power fluctuations of the FROSTI heating profile, produces displacement noise in the interferometer through photothermal and optomechanical couplings~\cite{Ballmer:2006, Willems:2011}. For an annular irradiance profile like that produced by the FROSTI, the dominant noise coupling is photothermal flexure, or ``bending,'' noise of the test mass. Flexure noise arises from thermoelastically-driven motion of the test mass' front surface relative to its center of gravity. Following Ref.~\cite{Willems:2011}, we estimate the flexure noise using an FEA model of a test mass in which the FROSTI heating profile shown in Figure~\ref{fig:frosti_concept} is applied with a harmonic perturbation. We find that the flexure noise contribution from each test mass can be expressed as
\begin{equation}
    \Delta z_{\mathrm{F}}(f) =  \left(5.21 \times 10^{-14}\;\mathrm{m}\right) \frac{10\;\mathrm{Hz}}{f} \, \frac{P}{1\;\mathrm{W}} \; \Xi(f)
    \label{eq:flexure}
\end{equation}
where $P$ is the average power applied by the FROSTI and $\Xi(f)$ is its relative amplitude noise spectral density (in units of $1/\sqrt{\mathrm{Hz}}$). We assume that the flexure noise contributions from the four test masses are uncorrelated and thus add in quadrature.

The FROSTI prototype's RIN is experimentally constrained using two infrared-sensitive ThorLabs PDAVJ5 photodetectors, positioned in front of the FROSTI reflector to maximize power incident on the sensors from a single heater element. Each photodetector signal is passed through a 2~kHz analog low-pass filter and sampled at 7.63~kHz by a Red Pitaya STEMlab 125-14 analog-to-digital converter (ADC). Following Ref.~\cite{Chou:2017}, we employ Welch’s modified periodogram method to compute the power spectral densities (PSDs) of the individual photodetector signals as well as their cross-spectral density (CSD), averaged over 39~hours of steady-state observing time. Since the electronic noises of the two photodetectors are largely uncorrelated, the time-averaged CSD can resolve small \textit{correlated} noise backgrounds to levels far below the noise floor of either photodetector individually. The CSD calculation procedure is described in further detail in Section~3 of the Supplemental Document.

\begin{figure}[t]
    \centering
    \includegraphics[width=1\linewidth]{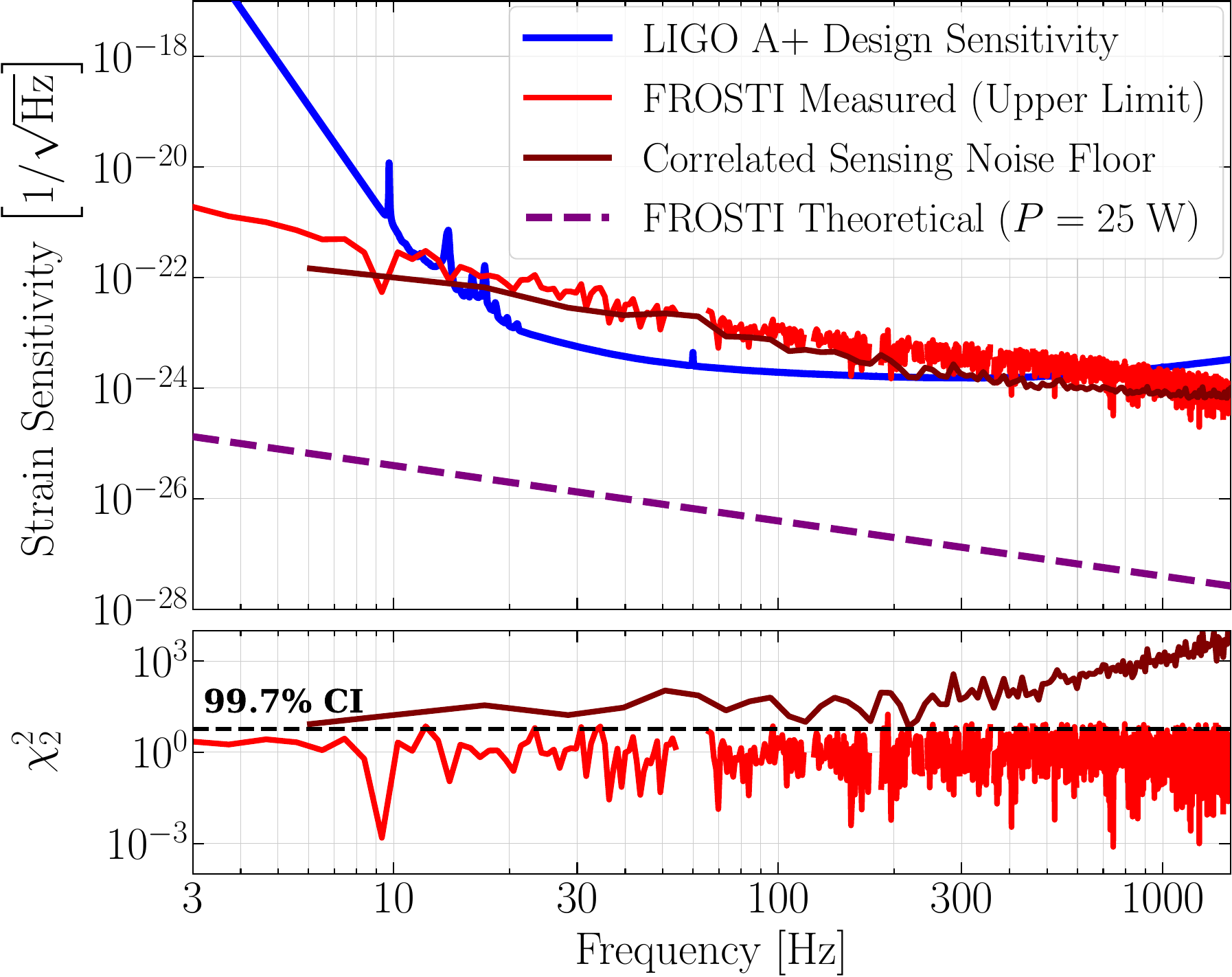}
    \caption{\textit{Top:} Measurement limits on the relative intensity noise of the FROSTI prototype (red), projected into units of strain sensitivity. Also shown is the LIGO A+ sensitivity target (blue), the correlated sensing noise floor of our two detectors (maroon), and the FROSTI's theoretical intensity noise for 25~W of delivered power (purple). \textit{Bottom:} Reduced chi-squared statistic of the cross-spectra-based measurements. Values greater than 5.9 (dashed line) indicate a detection of coherent intensity noise at the 99.7\% confidence interval (CI).}
    \label{fig:rin}
\end{figure}

Figure~\ref{fig:rin} shows the experimental limit placed by the CSD measurement on the RIN of the FROSTI prototype (red curve), projected into units of strain sensitivity. The measurement excludes the presence of coherent intensity fluctuations in the emission of the heater elements, to the indicated level of sensitivity, at 99.7\% confidence. The lower panel shows a reduced chi-squared statistic which quantifies the statistical significance of the curves in the upper panel. Its construction is detailed in Section~3 of the Supplemental Document. Values greater than 5.9 indicate a statistically significant detection of coherent noise at 99.7\% confidence (dashed black line), while values less than 5.9 indicate an exclusion. Also shown, for reference, is the LIGO~A+ sensitivity target (blue curve), the correlated electronic noise floor of the two ADC channels (maroon curve), and the FROSTI's theoretical intensity noise for 25~W of delivered power (purple curve). The presence of broadband correlated ADC noise imposes a systematic limit on the sensitivity achievable with this measurement apparatus. In future work, we will extend this measurement limit to lower levels by reducing the correlated sensing noise floor, through upgrades to the data acquisition infrastructure.

\subsubsection{Scattered Light Noise}
\label{sec:scatter}
Mounting a FROSTI actuator in front of a test mass introduces additional surfaces from which laser light scattered by the test mass can scatter \textit{back} into the main interferometer beam. Relative motion of the FROSTI, which is less seismically isolated than the suspended test mass, phase modulates the backscattered laser light. This phase noise couples directly to the apparent arm length~\cite{FlanaganScatteredLight}, with an amplitude spectral density given by
\begin{equation}
    \Delta z_{\rm PN}(f) = \sqrt{\frac{\epsilon}{2}} \; \xi(f)
    \label{eq:phase_noise}
\end{equation}
where $\epsilon$ is the fraction of incident laser power that backscatters and recombines with the main beam and $\xi(f)$ is the amplitude spectral density of the relative horizontal motion between the FROSTI and the test mass (in units of $\mathrm{m}/\sqrt{\mathrm{Hz}}$). The backscattered light also beats with the main laser field to produce fluctuations of the arm cavity power. These fluctuations coherently displace both test masses through radiation pressure~\cite{Fritschel:2013}, with an amplitude spectral density given by
\begin{equation}
    \Delta z_{\rm RP}(f) = \frac{4 \Gamma P_{\rm arm}}{\pi \lambda c M} \sqrt{2 \epsilon} \; \frac{\xi(f)}{f^2}
    \label{eq:radiation_pressure}
\end{equation}
where $P_{\mathrm{arm}}$ is the arm cavity power, $M = 40$~kg is the mass of the LIGO A+ test masses, $\lambda = 1064\;\mathrm{nm}$ is the laser wavelength, and $\Gamma = 14.3$ is the optical gain of the signal recycling cavity.

\begin{figure}[t]
    \centering
    \includegraphics[width=1\linewidth]{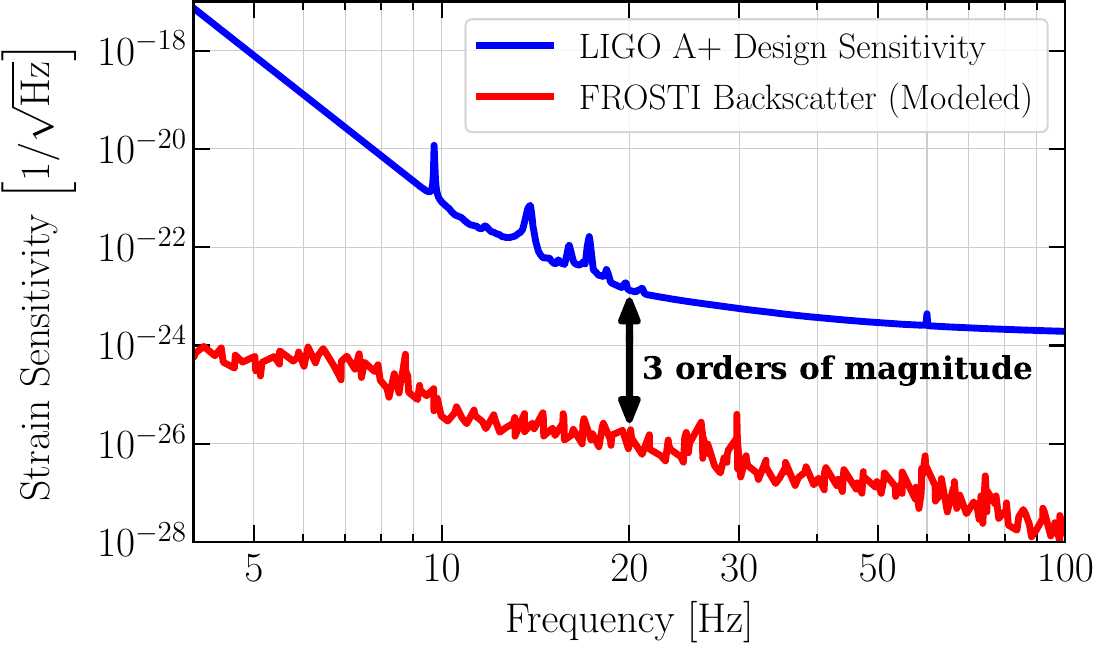}
    \caption{Total projected strain noise due to 1064-nm laser light backscattering from the FROSTI surfaces. At all frequencies, the modeled scattered light noise is found to lie at least three orders of magnitude below the target sensitivity of LIGO~A+.}
    \label{fig:scattered_light}
\end{figure}

Following the formalism in Ref.~\cite{FlanaganScatteredLight}, we estimate the fraction of incident laser power that backscatters from the FROSTI, finding that $\epsilon = 1.51 \times 10^{-23}$. This calculation is detailed in Section~4 of the Supplemental Document. We assume that the relative motion between the FROSTI and the test mass, $\xi(f)$, is the horizontal seismic noise spectrum of LIGO's seismic-isolation platform (BSC-ISI ST2)~\cite{Matichard:2015}, multiplied by a safety factor of 10 to account for additional controls-driven test mass motion. Using the above displacement noise couplings (eqs.~\ref{eq:phase_noise} and \ref{eq:radiation_pressure}), we project this noise into the differential arm length noise spectrum of the interferometer at nominal operating power ($P_{\rm arm} = 750\;\mathrm{kW}$), assuming that the backscatter noise contributions from each arm are uncorrelated and can be added in quadrature. Figure~\ref{fig:scattered_light} shows the total projected noise, in units of interferometer strain. We find that, at all frequencies, the scattered light noise lies at least three orders of magnitude below the target sensitivity of LIGO~A+.

\subsection{Ultra High Vacuum Compatibility}
\label{sec:vacuum}
\begin{figure}[t]
    \centering
    \includegraphics[width=1\linewidth]{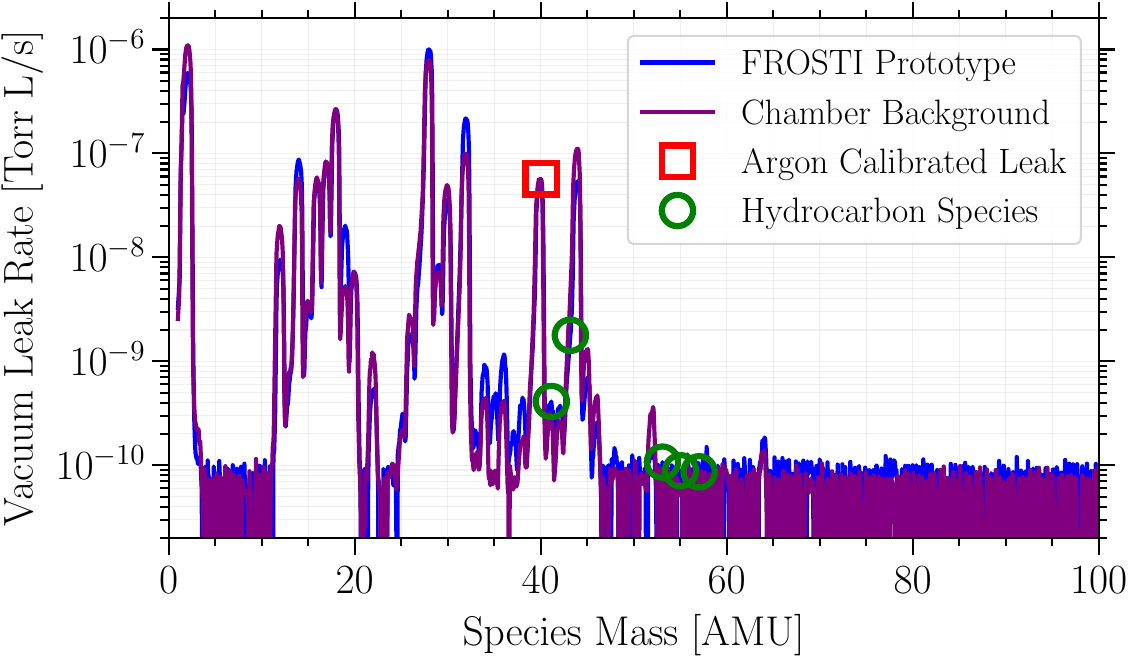}
    \caption{Measured outgassing rate spectrum of the FROSTI prototype operating at 625~K (blue), in comparison to the sensitivity limit imposed by the vacuum system's background (red). The red square indicates an injected argon calibration line. The green circles indicate the locations of hydrocarbon species.}
    \label{fig:rga_scan}
\end{figure}

To ensure compatibility with LIGO's ultra high vacuum (UHV) environment, the FROSTI actuator  must meet stringent outgassing standards~\cite{Coyne:2010}. Even trace amounts of hydrocarbons can severely damage the test masses under high laser power. We directly measure the vacuum outgassing rate of the FROSTI prototype, as a function of molecular species, using a residual gas analyzer (RGA) equipped with a calibrated argon leak. The argon leak releases gas into the system at a precisely known rate and can thus be used to calibrate the measured spectrum. The fully-assembled prototype was installed in a large vacuum chamber and operated at maximum power (with its heater elements at roughly 625~K) for two weeks. Figure~\ref{fig:rga_scan} shows the final measured outgassing rate spectrum of the prototype, in comparison to a reference measurement of the empty chamber indicating the background sensitivity limit. We find that the hydrocarbon signatures of AMUs 41, 43, 53, 55, and 57 are all consistent with LIGO's outgassing rate requirements~\cite{Coyne:2020}.

\section{Discussion}
\label{sec:discussion}
Our results demonstrate that the FROSTI prototype performs very closely to design expectations. In particular, the experimental measurements and analysis presented in \S\ref{sec:results} confirm all three of the key properties underlying the FROSTI concept:
\begin{enumerate}
    \item Our novel \textbf{nonimaging design technique} is an effective means of achieving wavefront actuation at higher spatial frequencies (20--50~$\mathrm{m^{-1}}$) in gravitational-wave detectors.
    \item The internal greybody radiation source is capable of achieving \textbf{high intensity stability}, while the reflector surfaces are expected to produce \textbf{negligible backscattered light noise}.
    \item It is possible to fabricate FROSTI actuators from entirely \textbf{ultra-high-vacuum-compatible materials}, which is critical for their compatibility with gravitational-wave detectors.
\end{enumerate}
Based on these findings, efforts are now underway to develop final FROSTI designs for the 40-kg test masses of LIGO~A+, incorporating practical lessons learned from the prototype testing.

\begin{figure}[t]
    \centering
    \includegraphics[width=1\linewidth, trim={8.5mm 1mm 10mm 3mm}, clip]{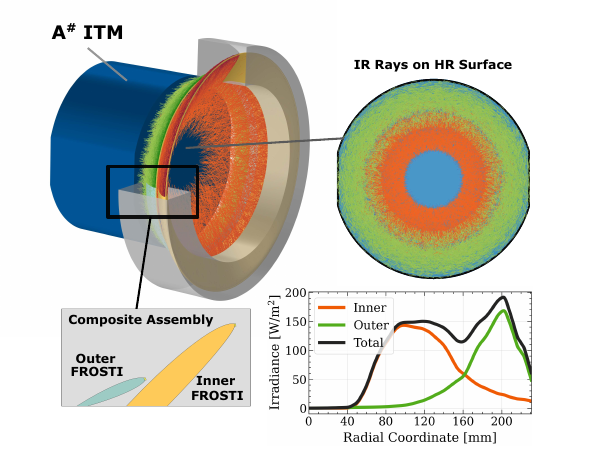}
    \caption{Concept for a next-generation FROSTI wavefront actuator, which superposes radiation from multiple heater rings (``Inner'' and ``Outer'') to create a complex front-surface heating profile. Such a profile, designed for a 100-kg ITM, can enable greater levels of laser power and squeezing in LIGO~$\rm A^{\#}$.}
    \label{fig:frosti_asharp}
\end{figure}

The impact of this work will also extend far beyond A+, as the new wavefront control capabilities demonstrated here lay the foundation for a critical piece of next-generation detector technology. Future extensions of the FROSTI concept will be integral to achieving the extreme power and squeezing targets of LIGO $\rm A^{\#}$ and, ultimately, Cosmic Explorer. Although the annular heating profile of the FROSTI prototype (see Figure~\ref{fig:frosti_concept}) does not deliver a sufficiently accurate wavefront correction for these future detectors in itself, preliminary modeling suggests that the irradiance profile shown in Figure~\ref{fig:frosti_asharp} (black curve), applied to the front surface of the ITM, could provide a sufficiently accurate correction to enable the $\rm A^{\#}$ power and squeezing targets. This heating pattern could be produced by nesting \textit{multiple} heater rings inside one composite assembly, as illustrated in Figure~\ref{fig:frosti_asharp} (left), with each heater ring enclosed by a separate nonimaging reflector that projects radiation onto a different radial zone of the test mass surface. This would allow the development of FROSTI actuators delivering more complex irradiance profiles, and hence more precisely-targeted wavefront corrections, than can be produced with a single heater ring alone. Thus, this work opens a key new research and development pathway towards realizing next-generation gravitational-wave observatories.

\begin{backmatter}

\bmsection{Funding}
National Science Foundation (NSF) (PHY-2110348, PHY-2409496).

\bmsection{Acknowledgments}
This material is based upon work supported by the National Science Foundation (NSF) under Award No. PHY-2110348. Additional support was provided by LIGO Laboratory under Advanced Detector Technology Research (ADTR) Initiative No. LIGO-M2200050. LIGO was constructed by the California Institute of Technology and Massachusetts Institute of Technology with funding from the NSF, and is operated by LIGO Laboratory under cooperative agreement PHY-1764464. This paper carries LIGO Document Number LIGO-P2500117.

\bmsection{Disclosures}
The authors declare no conflicts of interest.

\bmsection{Data availability}
Data underlying the results presented in this paper are not publicly available at this time but may be obtained from the authors upon reasonable request.

\bmsection{Supplemental document}
See Supplement 1 for supporting content. 

\end{backmatter}

\section*{}\vspace{-1\baselineskip} 
\bibliography{references}

\end{document}